\documentclass[epj]{svjour}
\usepackage{graphics}
\usepackage{amsmath}
\usepackage{amssymb}
\begin{document}
\title{Justification of Landau Hydrodynamic-Tube Model in Central Relativistic Heavy-Ion Collisions}
\titlerunning{Justification of Landau Hydrodynamic-Tube Model...}
\author{
U. U. Abdurakhmanov\inst{1} \and K. G. Gulamov\inst{1} \and V.V.Lugovoi\inst{1} \and V. Sh. Navotny\inst{1}
}                     
%
%
%
\authorrunning{U.U.Abdurakhmanov et al.}
\institute{Institute for Physics and Technology, Fizika-Solntse Research and Production Association,
Uzbek Academy of Sciences, Bodomzor yoli street 2, 100084 Tashkent, Uzbekistan}
\date{Received: date / Revised version: date}
%
\abstract{
The statistical event-by-event analysis of inelastic interactions of $^{16}$O and $^{32}$S nuclei in emulsion 
at 60 $A\,$GeV/c and 200 $A\,$GeV/c reveals the existence of groups of high multiplicity events belonging to
very central nuclear interactions with Gaussian pseudorapidity distributions for produced particles as
suggested by the original hydrodynamic-tube model. Characteristics of these events are presented. The
experimental observations are interpreted as a result of quark-gluon plasma formation in the course of
central nuclear interactions.
} 
\maketitle
\section{Introduction}
\label{intro}
Interest to the study of relativistic heavy ion collisions is caused by many reasons. After the
discovery of the quark-gluon plasma (QGP) at CERN and RHIC \cite{ref1,ref2,ref3,ref4,ref5} a particular attention is paid
to different issues related with the QGP formation and evolution in the course of nuclear
collisions at very high energies. The data reveal that central collisions of heavy ions at these
energies result initially in production of hadronic matter in the form of very hot compressed
and a nearly frictionless liquid (QGP) evolution of which produces the final state particles. The
data were analyzed in the framework of various theoretical approaches, including different,
sometimes very sophisticated, versions of the hydrodynamic model (for recent reviews see e.g.
\cite{ref6,ref7,ref8,ref9}). Considerations based on the hydrodynamic-type models were used not only for
discussion of more or less general properties of relativistic heavy ion collisions but also for the
analysis of more specific features like properties of hot compressed hadronic matter,
comparative yields of hadrons or two- and multiparticle correlations \cite{ref9,ref10,ref11,ref12} between particles
produced.

The shape of inclusive rapidity and pseudorapidity distributions of particles produced in
relativistic heavy ion collisions was discussed in many papers (see \cite{ref6} for a review). It was
shown \cite{ref13} that in a wide range of energies these distributions for symmetrical Au-Au and PbPb collisions 
may be described reasonably well by a Gaussian distribution which follows from
the original hydrodynamic model proposed by L.D.Landau \cite{ref14,ref15}. At the same time, it was
noticed \cite{ref6} that below $\sqrt{s_{NN}} < 10\, GeV$ the rapidity gap of the reaction for produced pions is
small, any created fireball with longitudinal expansion occupies its entire length with pions, so
that the agreement with a Gaussian shape may be fortuitous.

Of course, most of the data on multiparticle production in relativistic heavy ion collisions
analyzed so far belong to the inclusive and semi-inclusive reactions. At the same time some
important features of the production processes may be revealed more clearly if the
experimental data will be analyzed on the event-by-event basis. For example, if we analyze the
data in the framework of the hydrodynamic model the position of the central fireball on the 
longitudinal axis may vary because of geometrical reasons and no clear picture could be
revealed for the inclusive distributions. In this paper we analyze on the event-by-event basis
the experimental data on the shape of pseudorapidity distributions of relativistic singly charged
(shower) particles produced in central collisions of relativistic heavy ions in nuclear emulsion at
energies of the CERN SPS and the BNL AGS. More specifically we are looking at the possibility
that in central relativistic heavy ion collisions the pseudorapidity distributions of relativistic
singly charged particles in individual events follow the Gaussian shape as suggested in the
original hydrodynamic model \cite{ref14,ref15}.

\section{Experimental Data}
\label{sec:2}
The experimental data of the present paper were accumulated in the framework of the
EMU-01 collaboration \cite{ref16,ref17,ref18}. Emulsion stacks were irradiated by $^{16}$O nuclei 
at 60 $A\,$GeV/c and 200 $A\,$GeV/c, by $^{32}$S nuclei at 200 $A\,$GeV/c at the CERN SPS and 
by $^{197}$Au nuclei at 11.6 $A\,$GeV/c at the BNL AGS. In all cases the incident beams were parallel 
to the surface of emulsion plates, beams densities were approximately $5\cdot 10^3\, \text{nuclei/cm}^2$ 
with an admixture of foreign ions of no more than 2\%.
\begin{table*}
\centering
\caption{General characteristics of heavy-ion collisions considered in the present study}
\label{tab:1}       
\small
\begin{tabular}{|c|c|c|c|c|c|c|}
\hline
~~Projectile ~~&~~$E_0$, GeV~~&~~ $N_{ev}$ ~~&~~~  $n_s$  ~~~&~~~  $n_g$  ~~~&~~~  $n_b$  ~~~&~~ $N_h$~~ \\
\hline
$^{16}$O  &   60 &  884  & 42.5$\pm$1.5 &  5.7$\pm$0.4 & 4.5$\pm$0.2 & 10.2$\pm$0.9 \\
$^{16}$O  &  200 &  504  & 58.0$\pm$2.8 &  4.3$\pm$0.2 & 4.1$\pm$0.1 &  8.4$\pm$0.4 \\
$^{32}$S  &  200 &  884  & 80.3$\pm$3.3 &  4.7$\pm$0.3 & 3.9$\pm$0.2 &  8.6$\pm$0.7 \\
$^{197}$Au& 10.7 & 1057  & 80.7$\pm$2.5 &  5.9$\pm$0.2 & 3.6$\pm$0.1 &  9.5$\pm$0.3 \\
\hline
\end{tabular}
\end{table*}

For the analysis the events of inelastic incoherent interactions of incident nuclei in emulsion
were collected, the events of electromagnetic nature were excluded from consideration. In
accordance with emulsion technique, the secondary charged particles in events were divided
into different groups:

$black$ or $b$-particles, mainly consisting of protons from the target nuclei with momenta 
$ p \leq  0.2$ GeV/c and also heavier nuclear fragments;

$gray$ or $g$-particles corresponding to protons with the momenta $0.2 \leq p \leq 1$ GeV/c; they 
mainly consist of protons - fragments of the target nuclei, contribution of  slow pions does not 
exceed several percent. Black and gray particles may be combined into a group of strongly ionizing  
$h$-particles with a multiplicity $N_h = n_b + n_g$;

$shower$ or $s$-particles - singly charged particles with  speed $\beta \geq 0.7$ and ionization 
in emulsion $I < 1.4 I_0$ , where $I_0$ represents the minimal ionization on tracks of singly
charged relativistic particles. Shower particles consist mainly of produced particles
(pions) and the singly charged “spectator” fragments of the projectile nucleus. For the
emission angles of the latter ones we use the criterion \scalebox{1.14}{$ sin\theta_0 \leq \frac{0.2}{p_0}$}, 
where $p_0$ is the initial momentum per nucleon, so that particles with $\theta < \theta_0$ were excluded 
from sparticles, whose multiplicity is $n_s$ or simply $n$.

\textit{projectile fragments} - fast particles with charge $Z \geq 2$ and  ionization  $I/I_0 \approx 4$, 
not changing at long distances from the point of interaction in emulsion. These particles are not 
included into the number of \textit{b}- or \textit{g}-particles.

For all the above types of particles their multiplicity and the emission angles were determined. 
For the analysis of angular distributions of $s$-particles we use pseudorapidity: 
\begin{eqnarray}
\label{eq01}
\eta = - \ln \tan \frac{\theta}{2}
\end{eqnarray}
where $\theta$  is the emission  angle of  the $s$-particle. For pions pseudorapidity is related 
with true rapidity  by a simple equation:
\begin{eqnarray}
\label{eq02}
\sinh \eta = \frac{m_T}{p_T} \sinh y, \;\;\; \; \text{where} \;\; m_T^2 = m^2 + p_T^2.
\end{eqnarray}

Emulsion has complex composition consisting of groups of light (H,C,N,O) and heavy nuclei
(Br, Ag). Therefore selection of events in accordance with the conventional criterion $N_h \geq 8 $
corresponds to effective selection of interactions central with respect to heavy target nuclei (Br and Ag).

Some general information on the experimental data together with statistics is presented in
Table 1. More information is given in \cite{ref16,ref17,ref18,ref19}.

\section{The model}
\label{sec:3}

The hydrodynamic model of multiparticle production was originally developed for the head on nucleon-nucleon 
collisions at very high ( $> 1TeV$ ) energies of a projectile \cite{ref14}. In the c.m.
system colliding nucleons undergo a strong Lorentz contraction. At some instant the whole
initial energy is concentrated within a thin disk whose size coincides with the Lorentz contracted nucleon 
size. The disk is in rest in the c.m. system of colliding nucleons. The
hadronic matter within the disk has very high density and high temperature $T \gg \mu c^2$, where
$\mu$ is the pion mass, so that following modern concepts it consists of point-like quarks and
gluons, rather than usual hadrons. It is the quark-gluon plasma expanding according to the laws
of relativistic hydrodynamics of ideal fluid. While expanding, it becomes cooler. When the
temperature of hadronic matter reaches $T \approx \mu c^2$ , the plasma transforms into hadrons, mostly 
pions. The pseudorapidity distributions of produced particles in different reference systems
follow approximately a Gaussian (normal) shape, but only in the case of a very high multiplicity
does the pseudorapidity distribution in an individual event become a meaningful concept.

The model was generalized to the case of nucleon-nucleus collisions \cite{ref15}. In this case the 
projectile nucleon can cut out in the nucleus a tube whose cross section is equal to the cross
section of the nucleon and interacts only with this part of the target nucleus. The length of a
tube may vary in dependence of the geometry of an interaction. In contrast to the case of a
nucleon-nucleon collision, an intricate mechanism of compression of nuclear matter treated as
a continuous medium comes into play at the first stage of collision with a tube. After
compression, the one-dimensional (at the first stage) expansion of nuclear matter (the quark-gluon 
plasma, in modern terms) proceeds according to the laws of relativistic hydrodynamics of
ideal fluid. As in the case of a head-on nucleon-nucleon collision, the pseudorapidity
distribution of newly produced particles in a high-multiplicity nucleon-tube collision may be
approximated by a normal Gaussian distribution.

Hydrodynamic considerations were generalized also to the case of relativistic heavy ion
collisions (see, e.g. \cite{ref9,ref20,ref21,ref22,ref23}). As we understand, completely 
self-consistent and comprehensive hydrodynamic description of relativistic nucleus - nucleus collisions 
is not yet developed, but  ideas of relativistic hydrodynamics are used widely for both interpretation 
of different, sometimes very intriguing, aspects of the existing experimental data from the LHC and RHIC 
as well as for prediction of the trends of different experimental observables at these energies.
Moreover the hydrodynamic approach to multiparticle production was considerably enriched
and developed to incorporate new experimental findings in heavy ion collisions.

We are not discussing these issues. We are dealing with a rather simple old-fashioned
hydrodynamic approach considering a hot fireball representing by itself a compressed drop of
ideal hadronic fluid whose expansion and cooling leads to emission of final state particles with
Gaussian pseudorapidity distribution. The goal of the present paper is to study what the
experimental data tell us on this possibility.
\begin{figure*}
\centering
\resizebox{0.8\textwidth}{!}{
  \includegraphics{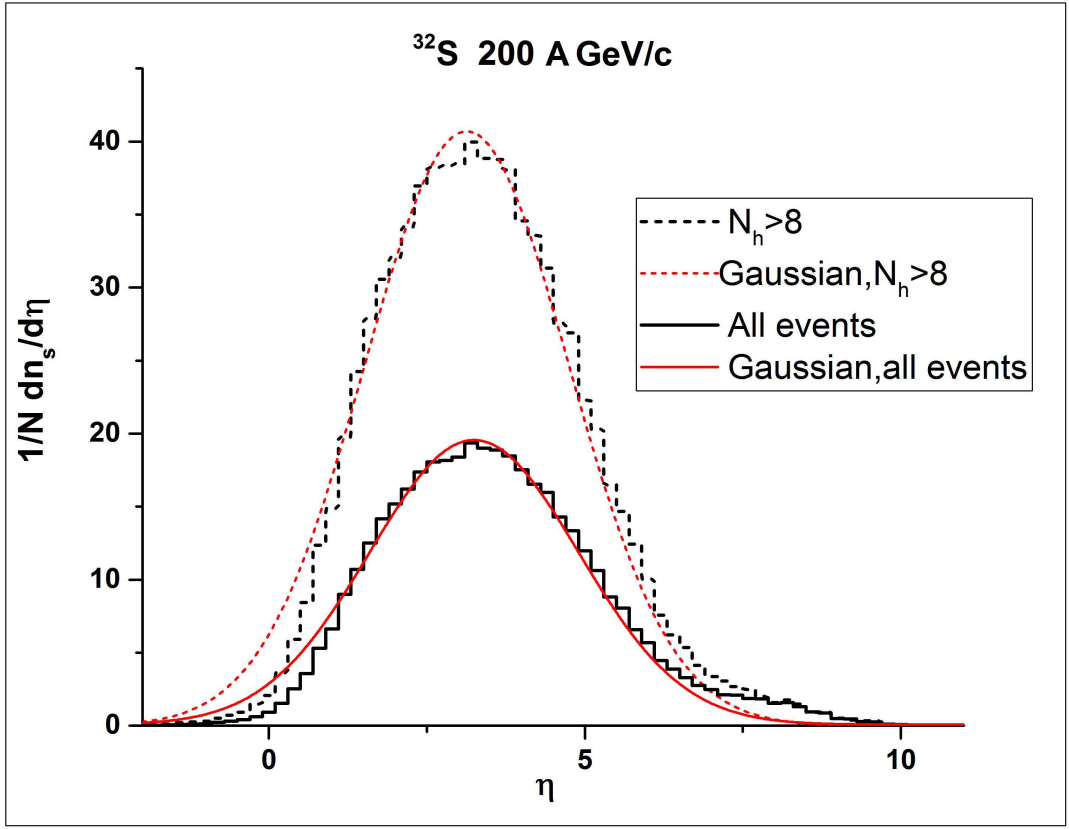}}
\caption{Pseudorapidity distributions for inclusive and semi-inclusive ($N_h \geq 8$) inelastic
interaction of $^{32}$S nuclei at 200 $A\,$GeV/c. Curves are Gaussian distributions.}
\label{fig:1}       
\end{figure*}

\section{Analysis of experimental data}
\label{sec:4}
In Figure 1 we present the experimental data on pseudorapidity distributions of relativistic
s-particles produced in interactions of $^{32}$S nuclei in emulsion. Separately we show also the data
for central collisions with respect to the heavy target nuclei ($N_h \geq 8$). The curves here 
represent the best fits to the data by Gaussian distributions. We see that in general the experimental 
data differ in shape from the Gaussian distributions, especially at small and high values of $\eta$. 
Even if we assume that in individual events the pseudorapidity distribution follows the Gaussian shape, 
the inclusive or semi-inclusive distributions, like in Figure 1, may decline from it because of
different reasons.

For the analysis of experimental data on the shape of pseudorapidity distributions of
relativistic particles in individual events we have applied the statistical approach described in
details in \cite{ref24}. We use the coefficient of skewness $g_1$, as a measure of asymmetry, and the
coefficient of excess $g_2$, as a measure of flattering, which represent parametrically invariant
quantities defined as (see Sect.15.8 in \cite{ref24})
\begin{eqnarray}
\label{eq03}
g_{1} &=& m_{3} m_{2}^{-3/2}  \;  , \;\;\;\; \notag \\
g_{2} &=& m_{4} m_{2}^{-2} - 3   \; , \;\;\;\;  \\ 
m_{k} &=& \frac{1}{n} \sum_{i=1}^{n} \left( \eta_{i} - \bar{\eta} \; \right)^{k} \; ,  \;\;\;\;    \bar{\eta} =  \frac{1}{n} \sum_{i=1}^{n} \eta_{i}  \;\;  \notag
\end{eqnarray}
where $m_k$ are the central moments of $\eta$ -distributions and $n = n_s$ stands here for the
multiplicity of $s$-particles in an event.

It follows from the mathematical statistics that if quantities $\eta_{1}$, $\eta_{2}$, ... , $\eta_{n}$ 
are independent of one another in events of a subensemble and obey Gaussian distributions, the distribution 
of these parametrically invariant quantities does not depend on the parameters of
the Gaussian distributions, and the number n of particles in the subensemble event 
uniquely determines the distribution of parametrically invariant quantities. In this 
case the mathematical expectation values and variances of $g_1$ and $g_2$ are 
as follows (see eq. (29.3.7) in \cite{ref24}):
\begin{eqnarray}
\label{eq04}
\nu_{g_{1}} (n) &=& 0   , \;\;\;  \sigma_{g_{1}}^{2} (n) =  6 (n-2) (n+1)^{-1} (n+3)^{-1} ,\notag\\ 
\nu_{g_{2}} (n) &=&  - 6 (n+1)^{-1}  , \;\;   \\
\sigma_{g_{2}}^{2}(n) &=&  24 n (n-2) (n-3) (n+1)^{-2} (n+3)^{-1} (n+5)^{-1}.\notag
\end{eqnarray}
We refer to the model described above, where the pseudorapidities obey a Gaussian
distribution, as the $G$ model.

From the mathematical point of view, our goal is to test the hypothesis that
pseudorapidities in the events with different and sufficiently large multiplicity $n$ 
are finite representative random samples with the volume n from the single infinite 
parent population (see Sect.13.3 in \cite{ref24}), in which pseudorapidities are 
distributed according to the Gaussian law. To test this hypothesis, we use the central 
limit theorem (see Sections 17.1-17.4 in \cite{ref24}), which asserts that the sum of 
a large number of independent and equally distributed so-called normalized random 
variables (see Sect.15.6 in \cite{ref24}) has a normal distribution in the limit. In
mathematical statistics, these normalized quantities are constructed from the random variable
and the mathematical expectation and variance obtained from these random variables (see
Sect.15.6 in \cite{ref24}). However, our goal is to test the hypothesis of the normality of 
pseudorapidity distribution in individual experimental events (that is, in the individual 
finite samples from an infinite parent population). Therefore, we construct a normalized 
random variable in a different way, namely: when constructing it for each individual event 
with a multiplicity of $n$, we calculate the quantities $g_1$ and $g_2$ (see eq.(3)), 
using the experimental values of the event pseudorapidities, and the variances and
mathematical expectations are determined by theoretical formulas (4) (see eq. (29.3.7) 
in \cite{ref24}) for a quantity with normal distribution.

Thus, if our hypothesis of normality is true (if the G-model is realized), then by our
construction, the normalized quantities $d_1$ and $d_2$ (see Sect.15.6 in \cite{ref24})
\begin{eqnarray}
\label{eq05}
d_{1} &=& \left[  g_{1} - \nu_{g_{1}} (n) \right] \;  \sigma_{g_{1}}^{-1} (n) \;,  \notag \\
d_{2} &=& \left[  g_{2} - \nu_{g_{2}} (n) \right] \;  \sigma_{g_{2}}^{-1} (n) \;
\end{eqnarray}
have dispersions equal to $1$ and mathematical expectations equal to $0$ both in the subensemble
of events (with the fixed number of particles $n$) and, consequently, in the ensemble of the
events (where $n$ can take any possible values).

Moreover, if the hypothesis of the normality of the pseudorapidtiy distribution is true, then,
according to the central limit theorem of mathematical statistics, for a sufficiently large number
$N$ of independent random samples (that is, the number of interaction events) the sums of these
independent and identically distributed normalized quantities
\begin{eqnarray}
\label{eq06}
\bar{d_{1}} \; \sqrt{N}  =  \frac{1}{\sqrt{N}}  \sum_{i=1}^{N} d_{1i} , \;\; \;\;
\bar{d_{2}} \; \sqrt{N}  =  \frac{1}{\sqrt{N}}  \sum_{i=1}^{N} d_{2i} \; 
\end{eqnarray}
should be less than 2 with the probability of 95\% (see Sections 17.1-17.4 in \cite{ref24})

If the hypothesis of normality is true, then the $G$-model is quite realistic and for a small $N$
we can use the asymptotic normality (see Sect.17.4 in  \cite{ref24}) of $g_1$ and $g_2$ in the 
subensemble of the events described by the $G$-model. Then the normalized quantities $d_1$ and $d_2$ are equally
distributed with parameters $0$ and $1$ in both the subensemble and in the ensemble of events
with the large enough $n^{min}$ to make the notion of distribution in individual event meaningful. In
this case (for the $G$-model) the sums (6) have the same restrictions.

In this paper, the sums (6) were calculated for interaction events with the multiplicity of
relativistic shower particles $n_s$ in the interval from $n_s^{min}$ to $n_s^{max}$. Calculations
were repeated for different intervals ($n_s^{min}$, $n_s^{max}$) with fixed $n_s^{max}$, 
whereas the value of $n_s^{min}$ was changing from some minimum value of $n_s$  to the maximum 
value of $n_s  = n_s^{max}$, which was defined from the experiment.

The procedures described were applied to all samples of our experimental data.

\begin{figure}
\centering
\resizebox{0.48\textwidth}{!}{
  \includegraphics{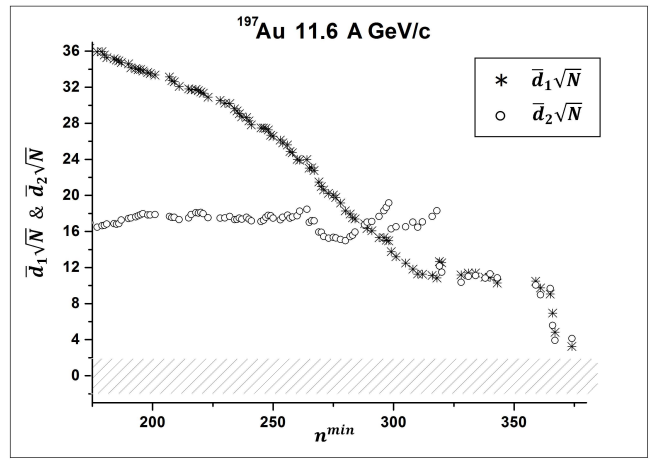}}
\caption{Dependence of parameters $\overline{d_1}\sqrt{N}$ and $\overline{d_2}\sqrt{N}$ on  
$n^{min}$ in interactions of gold nuclei in emulsion at 11.6 $A\,$GeV/c. The shaded area is 
the area where $ |\overline{d_1}\sqrt{N}| \; \& \; |\overline{d_2}\sqrt{N}| \; <\; 2$.}
\label{fig:2}       
\end{figure}
In Figure 2 we present the experimental data on values of the parameters 
$\overline{d_1}\sqrt{N}$ and $\overline{d_2}\sqrt{N}$ in dependence on the multiplicity $n^{min}$ 
in interactions of $^{107}Au$ nuclei in emulsion at 11.6 $A\,$GeV/c. We see that in general 
both $\overline{d_1}\sqrt{N}$ and $\overline{d_2}\sqrt{N}$ decrease in their absolute magnitude 
with $n^{min}$. At the same time it follows from this figure that in interactions induced by gold 
nuclei we have no events in which parameters $\overline{d_1}\sqrt{N}$ and $\overline{d_2}\sqrt{N}$ 
have values which are simultaneously less than $2$ in their absolute magnitudes. This means in 
the framework of our approach that there are no events in these interactions in which 
the pseudorapidity distributions of $s$-particles represent by themselves the statistical samplings 
from the Gaussian distributions. In general pseudorapidity distributions in these interactions 
do not follow the Gaussian shape.
\begin{figure}
\resizebox{0.48\textwidth}{!}{
  \includegraphics{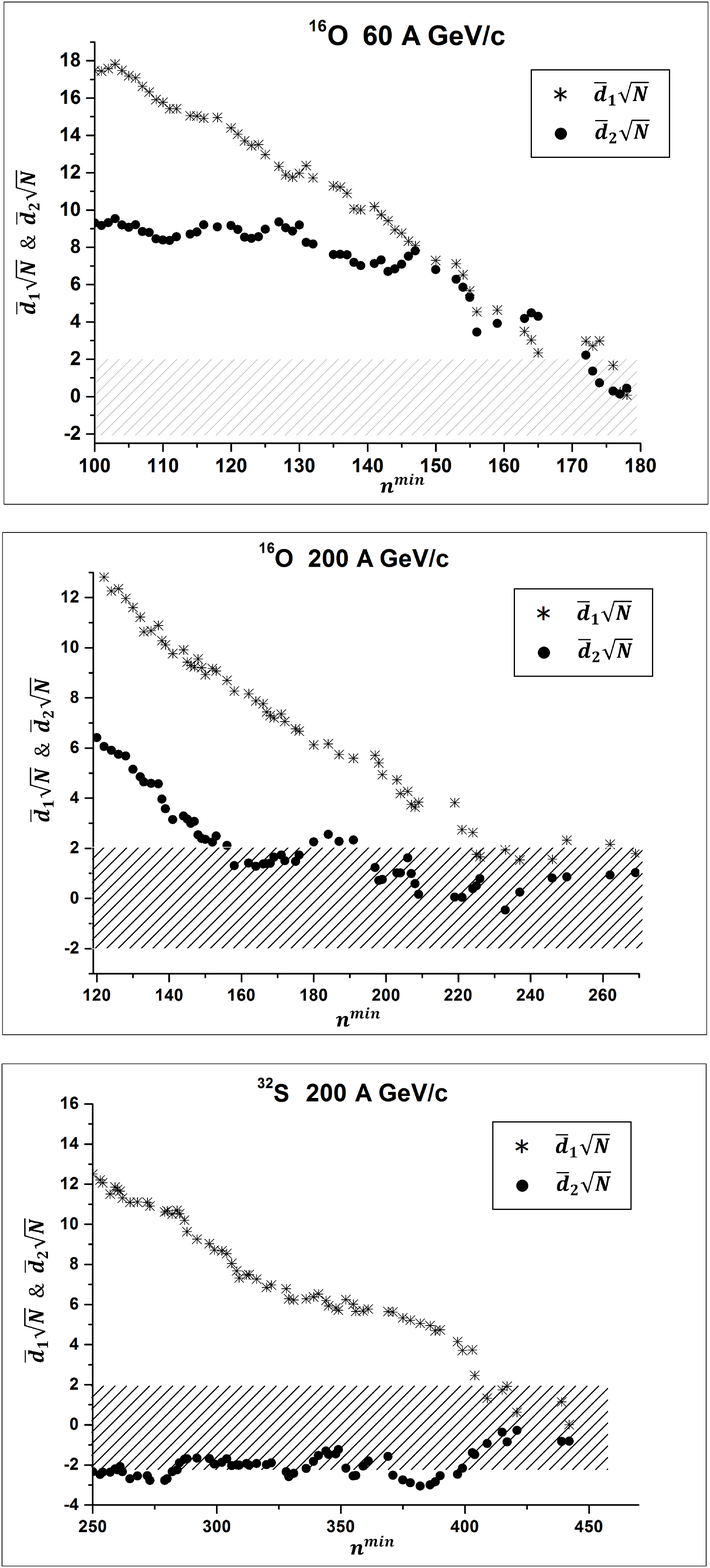}}
\caption{Dependence of parameters $\overline{d_1}\sqrt{N}$ and $\overline{d_2}\sqrt{N}$ on $n^{min}$ 
in interactions of oxygen nuclei in emulsion at 60 $A\,$GeV/c and 200 $A\,$GeV/c and in interactions 
of sulfur nuclei at 200 $A\,$GeV/c. The shaded area is the area where $ |\overline{d_1}\sqrt{N}| \; \& \; |\overline{d_2}\sqrt{N}| \; <\; 2$.}
\label{fig:3}       
\end{figure}

In interactions induced by $^{16}$O nuclei at 60 $A\,$GeV/c and 200 $A\,$GeV/c and by $^{32}$S nuclei 
at 200 $A\,$GeV/c we have also observed that the overwhelming majority of inelastic incoherent events
have pseudorapidity distributions of relativistic particles which do not follow the Gaussian
shape. Only in very small groups of very high multiplicity events we have found that both 
$\overline{d_1}\sqrt{N}$ and $\overline{d_2}\sqrt{N}$  have values which are less than $2$ in their 
absolute magnitudes and for these very events pseudorapidity distributions of $s$-particles represent 
by themselves statistical samplings from the Gaussian distributions. This fact is illustrated in 
Figure 3. Obviously pseudorapidity distributions in these events obey the Gaussian shape.

\begin{table}
\centering
\caption{The numbers and characteristics of events with the Gaussian pseudorapidity distributions}
\label{tab:2}       
\small
\begin{tabular}{|c|c|c|c|c|c|c|}
\hline
 &  &  &  &  &  &  \\
Projectile &$E_0,$ GeV & $N_{ev}$ & $n_s$ & $n_g$ & $n_b$ & $N_h$ \\
 &  &  &  &  &  &  \\
\hline
$^{16}$O  &   60 &  4  & 185.8 & 21.8  & 8.5 & 30.3 \\
$^{16}$O  &  200 &  9  & 247.0 & 18.8  & 8.8 & 27.6 \\
$^{32}$S  &  200 &  6  & 440.0 & 15.0  & 5.2 & 20.2 \\
\hline
\end{tabular}
\end{table}

The experimental data on the numbers of events in which pseudorapidity distributions of $s$-particles 
are Gaussian distributions together with average multiplicities in them are shown in
Table 2 for relativistic heavy-ion collisions analyzed in this paper. In Table 3 we show the
average values and dispersions of pseudorapidity distributions in these $9$ and $6$ individual
events found in interactions of $^{16}$O and $^{32}$S nuclei in emulsion at 200 $A\,$GeV/c. We see 
that average multiplicities of produced $s$-particles in these events are extremely high, exceeding 
(4-5) times average multiplicities in considered heavy ion collisions, so we are dealing with the
central heavy ion collisions. Comparing multiplicities of $h$-particles from Tables 1 and 2 we see
that these events belong to interactions of incident ions with heavy emulsion nuclei.
\begin{figure}
\centering
\resizebox{0.48\textwidth}{!}{
  \includegraphics{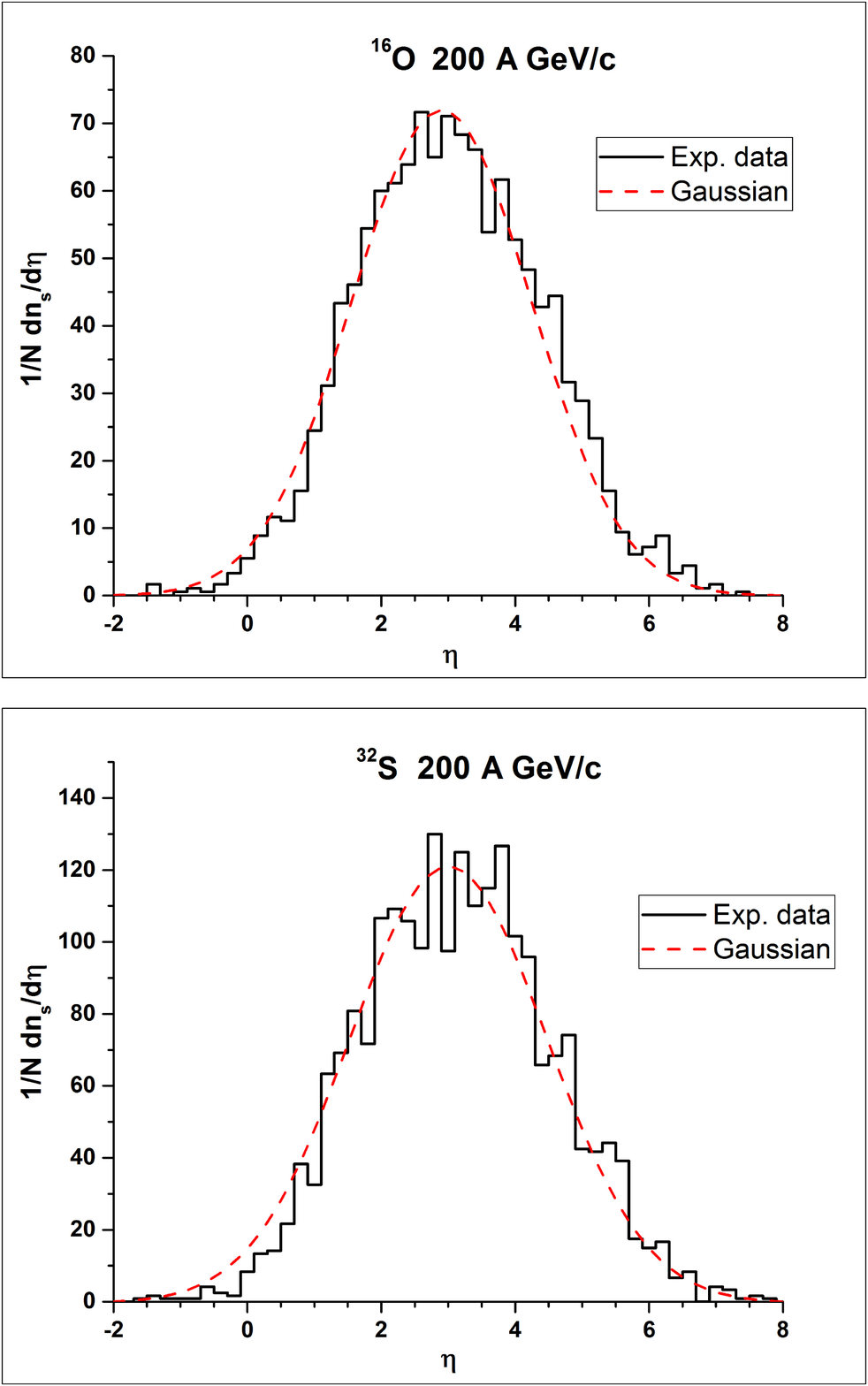}}
\caption{Pseudorapidity distributions in selected events for interactions of $^{16}$O and $^{32}$S 
nuclei at 200 $A\,$GeV/c. Curves are Gaussian distributions.}
\label{fig:4}       
\end{figure}

We see from the data of Table 3 that the centers and dispersions of pseudorapidity
distributions in events selected by our analysis do not fluctuate much, so that in Figure 4 we
show pseudorapidity distributions for all these events. Gaussian distributions describe them
well. In order to estimate the energy density for these events we can use approach suggested
by D.Bjorken \cite{ref25}. Of course, the exact values depend on some parameters, in particular 
on the radius of a volume, where the QCD transition could take a place. Having in mind our
experimental data on \scalebox{1.14}{$\frac{1}{N} \frac{dN}{d\eta}$} from Figure 4, our estimates may vary in the 
limits from $1 \text{ GeV/fm}^{3}$ to $ 16 \text{ GeV/fm}^{3}$, if the radius of the volume changes from 
$ 4 $ to $1\, \text{fm}$.
\begin{table}
\centering
\caption{Characteristics of pseudorapidity distributions in events selected by the analysis in interactions
of $^{16}$O and $^{32}$S nuclei in emulsion at 200 $A\,$GeV/c}
\label{tab:3}       
\small
\begin{tabular}{|c|c|c|c|c|c|}
\hline
\multicolumn{6}{|c|}{ }\\
\multicolumn{6}{|c|}{$^{16}$O, 200 $A\,$GeV/c}\\
\hline
No &~ $n_s$ ~~&~~ $\langle \eta \rangle $ ~&~ $\sigma(\eta)$ ~~&~~ $d_1\sqrt{N}$ ~~&~~ $d_2\sqrt{N}$ \\
\hline
1 & 232 &  2.91$\pm$0.09 &  1.41 &   0.58 &  -0.80 \\
2 & 245 &  2.98$\pm$0.09 &  1.44 &   0.55 &  -0.10 \\
3 & 225 &  2.79$\pm$0.09 &  1.37 &  -0.68 &   3.72 \\
4 & 249 &  3.02$\pm$0.09 &  1.36 &   1.36 &  -1.84 \\
5 & 232 &  2.97$\pm$0.10 &  1.48 &   0.27 &  -1.19 \\
6 & 275 &  3.02$\pm$0.07 &  1.17 &  -1.15 &   0.10 \\
7 & 261 &  3.10$\pm$0.08 &  1.35 &   0.91 &   0.09 \\
8 & 268 &  2.74$\pm$0.08 &  1.39 &   1.20 &   0.16 \\
9 & 236 &  2.82$\pm$0.09 &  1.35 &  -0.06 &  -1.14 \\
\hline
\multicolumn{6}{|c|}{ }\\
\multicolumn{6}{|c|}{$^{32}$S, 200 $A\,$GeV/c}\\
\hline
No &~ $n_s$ ~~&~~ $\langle \eta \rangle $ ~&~ $\sigma(\eta)$ ~~&~~ $d_1\sqrt{N}$ ~~&~~ $d_2\sqrt{N}$ \\
\hline
1 &  511 &  2.88$\pm$0.07 &  1.48 &  -0.65 &  -1.44 \\
2 &  420 &  2.96$\pm$0.07 &  1.36 &   0.04 &   0.86 \\
3 &  438 &  3.01$\pm$0.07 &  1.44 &   2.75 &  -1.22 \\
4 &  414 &  3.20$\pm$0.07 &  1.35 &  -0.53 &   0.68 \\
5 &  441 &  2.95$\pm$0.06 &  1.34 &   1.61 &  -0.35 \\
6 &  416 &  3.31$\pm$0.07 &  1.47 &   0.01 &  -0.82 \\
\hline
\end{tabular}
\end{table}

We have verified empirically the results of application of the statistical approach based on
parametrically invariant quantities $g_1$ and $g_2$ to the conditions of our experiments. In order to
do so we have used ensembles of Monte Carlo events generated following the
phenomenological model of independent emission of $s$-particles (IEM) \cite{ref26,ref27}. In the
framework of this model we assume that: (i) multiplicity ($n_s$) distributions of simulated events
reproduce the experimental distributions for the interactions considered; (ii) one-particle
pseudorapidity distributions of $s$-particles in each one of simulated subensembles of events
(within, for instance, the fixed range of $n_s$) reproduce the experimental distribution for the
same range of $n_s$; (iii) emission angles of $s$-particles in each one of simulated events are
statistically independent.

As an example in Figure 5 we show the values of parameters $\overline{d_1}\sqrt{N}$ and $\overline{d_2}\sqrt{N}$ 
in dependence on the multiplicity $n^{min}$ in Monte Carlo events generated in accordance with 
multiplicity and pseudorapidity distributions of $s$-particles for interactions of $^{32}$S nuclei 
in emulsion at  200 $A\,$GeV/c. We see that even at much higher statistics ($5000$ events) than the experimental 
ones Monte Carlo events do not reveal existence of any group of events with the Gaussian shape of
pseudorapidity distributions. We conclude from this figure that the probability of accidental
formation of the Gaussian pseudorapidity distributions in groups of individual events, not
recognizable by the present statistical approach, is negligibly small for the conditions of our
experimental data.
\begin{figure}
\centering
\resizebox{0.48\textwidth}{!}{
  \includegraphics{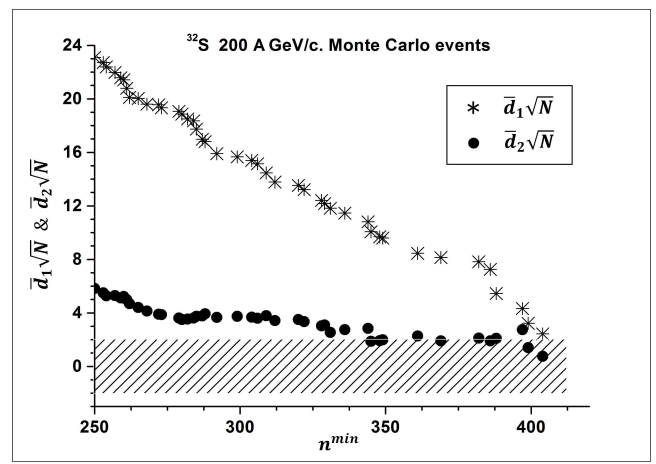}}
\caption{Dependence of parameters $\overline{d_1}\sqrt{N}$ and $\overline{d_2}\sqrt{N}$ on  $n^{min}$ 
in Monte Carlo events generated for interactions of sulfur nuclei in emulsion at 200 $A\,$GeV/c. The 
shaded area is the area where $ |\overline{d_1}\sqrt{N}| \; \& \; |\overline{d_2}\sqrt{N}| \; <\; 2$.}
\label{fig:5}       
\end{figure}

\section{Conclusions}
\label{sec:5}
In the present paper we have analyzed on the event-by-event basis the shape of
pseudorapidity distributions of produced particles in relativistic heavy ion collisions at CERN SPS
and BNL AGS energies. The goal was to search for events in which pseudorapidity distributions
of produced s-particles are Gaussian distributions. We have used for this purpose the statistical
method of parametrically invariant quantities [24]. Utilizing this approach to the experimental
data on interactions of $^{16}$O nuclei at 60 $A\,$GeV/c and 200 $A\,$GeV/c and interactions of 
$^{32}$S nuclei at 200 $A\,$GeV/c in emulsion we have discovered the existence of small groups of 
incoherent inelastic events with the Gaussian pseudorapidity distributions of produced particles. 
At the same time no events of this type were observed in interactions of gold $^{107}Au$ nuclei at 
lower incident energy – at 11.6 $A\,$GeV/c. Also we have found from results of Monte Carlo simulations
that the probability of accidental formation of the Gaussian pseudorapidity distributions in
high-multiplicity individual events is negligibly small for our experimental conditions.

The experimental data show that the multiplicity of produced particles in these events is
much higher than the average multiplicity in corresponding interactions and belongs to the high
end of multiplicity distribution, i.e. to the central heavy-ion collisions. For these events the
probability of formation is small enough (at the level of 1\% or less of the total statistics of
inelastic events) and most probably increases with the energy of an interaction.

The original hydrodynamic model is, to our best knowledge, the only model suggesting
some certain shape for pseudorapidity distributions of produced particles - the Gaussian
distribution. Simplicity of the model probably is one of the reasons why the original
hydrodynamic model is considered to be a “wildly extremal proposal” \cite{ref6}. One can note, of
course, that the model was introduced to describe only few general characteristics of
multiparticle production processes, pseudorapidity distribution of charged particles being one
of the simplest characteristics of the production process. So, the scope of the model was rather
narrow. To describe more complex features and characteristics of the production process, like
observed in modern experiments, it is necessary to go beyond and to exploit more advanced
versions of the hydrodynamic model. From this point of view modern versions probably are
more plausible but less certain in predictions.

The experimental observations of the present paper encourage us to interpret the
existence of events with the Gaussian pseudorapidity distributions of produced particles in
central relativistic heavy-ion collisions as a result of formation in the course of an interaction of
a droplet of hadronic matter - the quark-gluon plasma, i.e. the primordial high density state,
whose expansion and cooling leads to its decay with production of final state particles. This
interpretation may be supported by following considerations.

Calculations in the framework of lattice QCD show \cite{ref28,ref29} that at the energy densities
exceeding a critical value of about $1$ to $1.5 \text{ GeV per fm}^{3}$ achievable at incident energies 
of about $\sqrt{s_{NN}} \gtrapprox 5\text{ GeV} $, the hadronic phase of matter disappears, giving rise to 
the primordial high density state (QGP) whose evolution is governed by the elementary interactions 
of quarks and gluons. One can note that top SPS energies exceed by far these incident energies and
realization of such a phase transition in heavy-ion collisions at these energies was confirmed
experimentally by many characteristic signals (see, e.g.\cite{ref30}).
 
Of course, realization of the phase transition cannot be taken for granted in all relativistic
heavy ion collisions at these energies, even central ones. It is a rather rare and random
phenomenon and at the energies considered fluctuations in the energy density could play an
important role so that the produced primordial QCD objects may vary in some initial
characteristics, in the volume, for example. Therefore it was recommended at the beginning of
the QGP age to search for these objects experimentally on the event-by-event basis \cite{ref31}.
Evolution of these objects is probably reflected by the original hydrodynamic model and may
lead to the Gaussian distributions of final state particles in pseudorapidities. Therefore we
believe that it is important to study and to confirm this possibility in other experiments as well.

We are grateful to all members of the EMU-01 collaboration, and especially to professor
Ingvar Otterlund, for the joy to work together and for the excellent quality of the data.

%
%

\end{document}